\documentclass[english,aps,pra,showpacs]{revtex4}
\usepackage{graphicx}
\usepackage{babel}

\begin{document}

\title{Quantum Hall fractions for spinless Bosons}

\author{N.~Regnault}

\author{Th.~Jolicoeur}
\email{Nicolas.Regnault@lpa.ens.fr,
Thierry.Jolicoeur@lpa.ens.fr}

 \affiliation{Laboratoire Pierre Aigrain, ENS,
D\'epartement de Physique, 24, rue Lhomond, 75005 Paris, France}

%%%%%%%%%%%%%%%%%%%%%%%%%%%%%%%%%%%%%%%%%%%%%%%%%%%%%%%%%%%%%%%%%%%%%%%%%%%%%%%%%

\begin{abstract}
We study the Quantum Hall phases that appear in the fast
rotation limit for Bose-Einstein condensates of spinless bosonic atoms.
We use exact diagonalization in a spherical geometry to obtain
low-lying states of a small number of bosons as a function of
the angular momentum. This allows to understand or guess the physics
at a given filling fraction $\nu$, ratio of the number of bosons to the
number of vortices. This is also the filling factor of the lowest
Landau level. In addition to the well-known Bose Laughlin state
at $\nu =1/2$ we give evidence for the Jain principal
sequence of incompressible states at $\nu =p/(p\pm 1)$ for a
few values of $p$. There is a collective mode in these states
whose phenomenology is in agreement with standard arguments coming
e.g. from the composite fermion picture.
At filling factor one, the potential Fermi sea of composite fermions
is replaced by a paired state, the Moore-Read state. This is
most clearly seen from the half-flux nature of elementary excitations.
We find that the hierarchy picture does not extend up to the point of
transition towards a vortex lattice. While we cannot conclude, we investigate
the clustered Read-Rezayi states and show evidence for incompressible states
at the expected ratio of flux vs number of Bose particles.
\end{abstract}


\maketitle

%%%%%%%%%%%%%%%%%%%%%%%%%%%%%%%%%%%%%%%%%%%%%%%%%%%%%%%%%%%%%%%%%%%%%%%%%%%%%%%%

%introduction

\section{Introduction}

Right from the beginning of the theory of the fractional Quantum
Hall effect (FQHE)~\cite{Laughlin83}, it has been clear that the appearance of
incompressible liquids is due to the interplay of repulsive
interactions and the peculiar dynamics of the lowest-lying Landau
level (LLL) but is not tied in any fundamental way to the
statistics of the intervening particles. For example, in the
hierarchy scheme described by Haldane in ref.~\cite{Haldane83} the
quasiparticles are described by the Bose statistics and as a
consequence they condense into Laughlin states with an \emph{even}
power of the relative separation. While theoretically important,
this point has not received much attention till the advent of
trapped ultra-cold atomic vapors that can be set in
rotation~\cite{Matthews99,Madison00,Haljan01,Abo01,Rosenbusch02}.
It is now feasible to gather a large number of bosonic atoms in a
trap (magneto-optical or purely optical)
and to manipulate many parameters in a
controlled way. Notably one can create rotating Bose-Einstein
condensates. They display the various phenomena expected for a
rotating superfluid. When the velocity is large enough vortices
are nucleated and eventually when their number grows they form the
celebrated Abrikosov triangular lattice. This lattice can be
considered as an elastic medium which has peculiar vibration
modes. It has been pointed out in ref.~\cite{Wilkin98} that for a
special relation between the total angular momentum of the system
and its number of particles, the ground state for bosons with a
$\delta$ interaction is exactly a Laughlin wavefunction
corresponding to a filling $\nu =1/2$ of the LLL when the system
is in a two-dimensional (2D) regime. So, for large angular
momentum the physics of the vortex lattice should be replaced by
that of the FQHE for bosons~\cite{Cooper99,Wilkin00,Cooper01}.
If we consider a trap which is
strongly confining along one direction so that the motion is
effectively 2D, then in the perpendicular plane particles will
feel a (in general) harmonic restoring force. If the system is
rotating there will be an additional centrifugal force
counteracting the restoring force. If the rotation frequency
equals the characteristic frequency of the harmonic trap, then we
are left with only the Coriolis force which is formally equivalent
to a magnetic field. The bosons will then be exactly in the
conditions expected for the appearance of the FQHE. The extensive
degeneracy of the one-particle problem, i.e. the degeneracy of the
LLL will be lifted by interactions. For spinless bosons at
ultracold temperatures, the scattering takes place only in the $s$
wave and so there is an effective local $\delta$ interaction
between particles.

So far, the FQHE has only been observed experimentally for
fermions with Coulomb interactions (including spin and layer index
in bilayer structures). It is thus of prime importance to
investigate the novelties that arise in the Bose setting. It is
natural to expect a whole hierarchy of incompressible liquids with
the accompanying quasiparticles. The fractional charge means here
excess or missing density. In addition to the Laughlin state at
$\nu =1/2$ which is exact, the quasiholes created from this parent
liquid are also exact zero-energy eigenstates of the Hamiltonian
of this system. It has been suggested that the fractional
statistics of these entities may be experimentally measured by
laser manipulations~\cite{Paredes01}. There is evidence from exact
diagonalizations
of systems with a small number of particles for a prominent
sequence of fractions $\nu =n/(n\pm 1)$, akin to the so-called
principal sequence $\nu =n/(2n\pm 1)$ observed in fermion systems.
While this is closely related to the physics of the fermion
systems, there is also evidence~\cite{Cooper01}
for more exotic states intervening
in the filling range between $\nu \approx 1$ and the melting
transition of the vortex lattice which may take place for $\nu
\approx 6$. For the filling $\nu =1$, it has been suggested that
the ground state is closely related to the so-called Pfaffian state
of Moore and Read~\cite{Moore91} proposed
to describe fermions at $\nu =5/2$. One can describe this Pfaffian
state by pairing of fictitious composite fermions. The pairing has
two signatures that can be seen in numerical experiments~: there
is a special shift in the relation between the flux and the number
of particles in the spherical geometry and also quasiparticles are
created in pairs by adding one extra flux quantum onto the
fiducial ground state. For fillings $\nu >1$, there is some
evidence from exact diagonalizations in the torus geometry that
the incompressibility occurs through the Read-Rezayi
states~\cite{ReadRezayi}. These
states are generalizations of the Pfaffian and involve the
clustering of $k$ composite particles at filling $\nu =k/2$ in
their simplest version. They seem to fill the gap between the
Pfaffian state at $\nu =1$ and the phase transition to the vortex
lattice. The critical filling for this transition has been
evaluated from exact diagonalization~: the ordering is identified
by the appearance in the spectrum of low-lying states with quantum
numbers appropriate to the breaking of translational invariance in
the torus geometry~\cite{Cooper01}.
A wealth of new quantum Hall states is also predicted for bosons
with spin~\cite{Reijnders02,Reijnders03}. Some of them are generalizations
of the clustered Read-Rezayi states.
For spinless bosons, there is so far no rigorous proof that exotic
paired states are realized in the realistic case of pure s-wave
scattering. However if one gets close to a so-called Feshbach
resonance the resulting mixture of atoms and molecules may admit such
exotic states as rigorous ground states~\cite{Cooper03}.

In this paper, we give the detailed results of a study of the FQHE
for spinless bosons by exact diagonalization mainly in the spherical
geometry. By studying the low-lying levels of small systems, we
identify candidates for series of fractions displaying
incompressibility. We give evidence for the existence of a
Pfaffian-like ground state at $\nu =1$. We have also studied
larger fillings and have identified potential candidates for the
Read-Rezayi states that have so far only been observed in the
torus geometry. We examine ground state degeneracy as well as the
peculiar charged excitations. Some of our results were announced
in a Letter~\cite{Regnault03}.

In section II, we give some details of the formalism used to treat
the Bose problem on the disk and on the sphere. In section III, we
give evidence for the principal sequence of fractions. Section IV
is devoted to the filling $\nu =1$ and its relation to the
Pfaffian. Section V contains the results about fillings larger
than one. Finally section VI contains our conclusions.

%%%%%%%%%%%%%%%%%%%%%%%%%%%%%%%%%%%%%%%%%%%%%%%%%%%%%%%%%%%%%%%%%%%%%%%%%%%%%%%%
\section{Formalism~: from the disk to the sphere}
In the rotating frame, the Hamiltonian describing N interacting
bosons of mass $m$ is given by~:
\begin{equation}\label{Ham1}
  \mathcal{H} = \sum^{N}_{i=1}\frac{1}{2m}(\mathbf{p}_{i}-m\omega
    \mathbf{\hat{z}}\times \mathbf{r}_{i})^{2}
    +\frac{1}{2}m(\omega_{0}^{2}-\omega^{2})(x_{i}^{2}+y_{i}^{2}) \\
   +\frac{1}{2}m\omega_{z}^{2}z_{i}^{2}+\sum_{i<j}^{N}V(\mathbf{r}_{i}-\mathbf{r}_{j}).
\end{equation}
Here, the angular velocity is $\omega$ along the
$\mathbf{\hat{z}}$ axis, the $xy$ trap frequency is $\omega_0$ and
the characteristic frequency of the trapping $z$ potential is $\omega_z$.
We have
a 2D regime when the confinement along $z$ is strong enough to
freeze the motion in its ground state. When we tune the rotation
close to criticality $\omega$ $\approx$ $\omega_0$, we are dealing
with the problem of bosons in an effective field
$\mathbf{B}=(2m\omega/e)\mathbf{\hat{z}}$. There is a
corresponding magnetic length $\ell=\sqrt{\hbar/(2m\omega)}$. The
one-body problem displays the Landau levels with a degeneracy
given by $1/2\pi\ell^2$ state per unit area. In the language of
the quantum Hall effect, it is convenient to define the filling
factor $\nu =n\times 2\pi\ell^2$ where $n$ is the areal density of
bosons. In the language of rotating traps the filling factor $\nu$
is the ratio of the number of bosons to the average number of
vortices. For large number of particles and vortices $\nu \approx
N^2/2L_z$ where $L_z$ is the angular momentum along the
$\mathbf{\hat{z}}$ axis. When we are exactly at the point $\omega$
= $\omega_0$, the system has a degeneracy which is lifted only by
the interactions. In the case of ultracold bosons, the scattering
between atoms takes place only in the $s$-wave. As a consequence,
it is an excellent approximation to describe interactions by an
instantaneous potential which is a delta function~:
\begin{equation}\label{Potential}
    V(\mathbf{r})=\frac{4\pi\hbar^2
    a_s}{m} \, \delta^{(3)}(\mathbf{r}),
\end{equation}
where $a_s$ is the $s$-wave scattering length.
We consider the repulsive case $a_s >0$.
If we are in a 2D
regime, the wavefunction along z axis will be the ground state of
the harmonic oscillator and thus the effective potential felt in
the plane is~:
\begin{equation}\label{Pot2D}
 V^{(2D)}(\mathbf{r})\equiv \sqrt{32\pi}\,\, \hbar\omega \,\frac{a_s\ell^{2}}{\ell_z}
 \,\, \delta^{(2)}(\mathbf{r}).
\end{equation}
In this equation, $\ell_z$ is the characteristic length of the
z-axis oscillator~: $\ell_z = \sqrt{\hbar/m\omega_z}$. Since the
kinetic energy is totally quenched, all characteristic energies
will be given by pure numbers times the energy scale given by the
dimensionfull coefficient appearing in $V^{(2D)}(\mathbf{r}) $~:
$g=\sqrt{32\pi}\,\hbar\omega \, (a_s/\ell_z)$. The separation of the
successive Landau levels is given by $\hbar \omega_c =2\hbar\omega$.
Comparing this cyclotron energy to the scale $g$ set by interactions,
we see that Landau level mixing will be small if $a_s/\ell_z$ is
small enough.

The Hamiltonian in Eq.(\ref{Ham1}) is that of charged particles in
the symmetric gauge with a vector potential
$\mathbf{A}(\mathbf{r})=\frac{1}{2}\mathbf{B}\times\mathbf{r}$. We
will from now on concentrate on the lowest Landau level only. The
eigenstates of the one-body problem are then simply given by
functions of the complex coordinate $z=x+iy$~:
\begin{equation}\label{OneBodyEigen}
    \phi_m (z)=\frac{1}{\sqrt{\ell\sqrt{\pi}}}\,\, z^m \,
    \mathrm{e}^{-|z|^2/4\ell^2}.
\end{equation}
The \textit{positive} integer $m$ is the angular momentum of the
state~: $L_z =m\hbar$. Due to the very special dynamics of the
LLL, it is easy to solve the two-body problem for an arbitrary
interaction between particles. The eigenstate of the relative
particle is taken to be of the form Eq.(\ref{OneBodyEigen}) and
then the associated eigenenergy is $\hbar\omega_c +V_m$ where $V_m
= \langle \phi_m|V|\phi_m\rangle$. The coefficient $V_m$ is the
interaction energy of two particles with relative angular momentum
$m$. These coefficients called pseudopotentials are a convenient
way to parameterize interactions~\cite{Haldane85}. In the case of the pure delta
interaction, only $V_0$ is different from zero. In the case of the
Coulomb interaction all pseudopotentials are non zero and only the
ones with even $m$ contribute to the interaction between bosons.

In order to be entirely in the LLL, an arbitrary many-body state
$\Psi$ should be of the following from~:
\begin{equation}\label{LLL}
    \Psi (x_1,y_1,\dots , x_N,y_N)=P(z_1,\dots ,z_N)\prod_{k=1}^{N}
    \mathrm{e}^{-|z_k|^2/4\ell^2},
\end{equation}
where $P$ should be an analytic function of each of the $z_i$
complex coordinates. The Bose statistics requires $P$ to be
symmetric.

The many-body problem so defined is finite-dimensional even in the
unbounded plane once we fix the total angular momentum $L_z^{tot}$
since all the states in Eq.(\ref{OneBodyEigen}) have positive angular
momentum. So for a small number of bosons one may diagonalize
exactly the Hamiltonian in the disk geometry in sectors of fixed
$L_z$. The ground state energy is then a decreasing
function of $L_z^{tot}$ since the particles interact only for zero
relative angular momentum. Zero ground state energy is reached for
$L_z=N(N-1)$ by the celebrated Laughlin-Jastrow wavefunction~:
\begin{equation}\label{LJ}
    \Psi_{LJ}^{(2)} = \prod_{i<j} (z_{i}-z_{j})^2 \,\,\prod_{k=1}^{N}
    \mathrm{e}^{-|z_k|^2/4\ell^2}.
\end{equation}
It is only starting from this value of $L_z$ that all particles
can completely avoid each other. For the special condition
$L_z=N(N-1)$ this is in fact the only zero-energy eigenstate.
For larger values of the angular momentum, there are more and more
zero-energy states obtained by multiplying $\Psi_{LJ}^{(2)}$ by an
arbitrary symmetric polynomial $Q$ of the $z_{i}$ coordinates~:
\begin{equation}\label{Zero}
    \Psi = Q(z_1,\dots ,z_N)\times \Psi_{LJ}^{(2)}
\end{equation}
It is well known from the theory of the fermionic FQHE that when
the degree of $Q$ is small enough, the states Eq.(\ref{Zero})
describe edge excitations of the Laughlin droplet~\cite{Cazalilla03}.
When the degree
of the polynomial is of order $N$, one generates quasiholes. the
polynomial
\begin{equation}\label{QH}
 Q(z_1,\dots ,z_N) = \prod_{i}(z_{i}-Z_0),
\end{equation}
will generate a quasihole located at $Z_0$. The delta function
interaction for which the Laughlin wavefunction is an exact ground
state  has also the properties that the quasiholes are exact,
zero-energy eigenstates. This however does not extend to the
quasielectrons as we will see below. When $Q$ has a degree of
order $N^2$, it will finally change the filling factor of the LLL.

\begin{figure}[!htbp]
\begin{center}\includegraphics[width=3.25in,
  keepaspectratio]{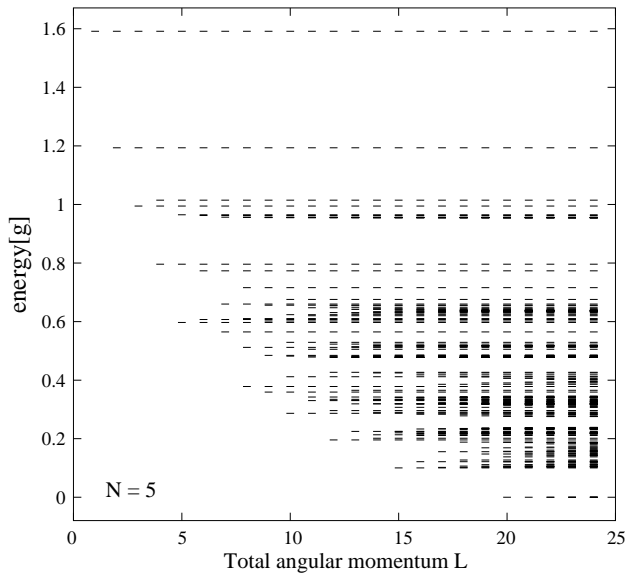}
\end{center}
\caption{Energy spectrum for N=5 bosons.
Energies are in units of $g$ and the horizontal axis
is total angular momentum. The Laughlin state appears at zero energy for
L=20.}
\label{disk}
\end{figure}

A sample spectrum is displayed in Fig.(\ref{disk}) for N=5 bosons. Each
state at the left boundary of the spectrum, i.e. the one with
smallest $L_z^{tot}$ at fixed energy, generates a family of
daughter states with increasing angular momentum. For zero energy,
the family is the one we described above in Eq.(\ref{Zero}). It is
clear that when performing exact diagonalization in a sector with
fixed $L_z$ we have to cope with all the descendants of the states
with smaller angular momentum~\cite{Trugman85,Bertsch99,Viefers00}.
A more appealing
scheme for such numerical studies is to use the spherical geometry
in which levels can be classified by the full rotation group. It
was shown by Haldane and Rezayi~\cite{Haldane85} that the
essential phenomena of the FQHE can
be understood from such calculations even for a relatively small
number of fermions.

If the bosons reside on a sphere then we need a magnetic monopole
at the center of the sphere in order to produce a uniform magnetic
field on its surface. If $R$ is the radius of the sphere, then
Dirac quantization condition requires that the total flux $4\pi
R^2 B$ must be an integral multiple $2S$ of the flux quantum
$h/e$. Thus the magnetic field is proportional to $S$~: $B=\hbar
S/eR^2$. The one-body problem can then be solved in polar
coordinates~\cite{Tamm31,Wu76}~: the LLL has degeneracy $2S+1$
and is spanned by the
(unnormalized) basis functions~:
\begin{equation}\label{LLLBase}
    u^{S+M}v^{S-M},\,\,\mathrm{where}\,\, M=-S,\dots ,+S,
\end{equation}
and $u$ and $v$ are the spinors in spherical coordinates~:
\begin{equation}\label{Spinors}
    u=\cos (\theta /2)\mathrm{e}^{i\varphi/2}, \, \,
    v=\sin (\theta /2)\mathrm{e}^{-i\varphi/2}.
\end{equation}
The sphere and the unbounded disk can be related by a
stereographic projection~\cite{Fano86} that allows to convert
explicit
wavefunctions from one geometry to the other. The previous
Laughlin wavefunction Eq.(\ref{LJ}) becomes on the sphere~:
\begin{equation}\label{SLJ}
    \Psi^{(2)}=\prod_{i<j}(u_i v_j-u_j v_i)^2 .
\end{equation}
In order to reside entirely in the LLL, the total degree in $u$
and $v$ of an arbitrary wavefunction for N particles should be
$2NS$. The above function $\Psi^{(2)}$ has total degree $2N(N-1)$
so it will be an acceptable state only for the precise matching
between flux and the number of particles $2S=2(N-1)$. When both
$N$ and $2S$ are large, this is what we expect from the filling
factor $\nu =1/2$ of the function $\Psi^{(2)}$. However for all
finite values there is a nontrivial shift between finite system
matching condition and its thermodynamic limit. This shift has a
topological origin and can be used to discriminate between
candidates for incompressible states. It will be useful in our
study of the Pfaffian state described in section IV.

On the sphere, total angular momentum is conserved and the
many-body eigenstates can be classified by their $L_{tot}$ value.
An incompressible quantum Hall fluid which breaks no symmetry is
expected to appear as an isolated $L=0$ singlet separated by a
sizable gap from the excited states. By comparison, broken
symmetry states appear as a set of quasi-degenerate states. The
gap above the isolated singlet is the hallmark of
incompressibility, the most salient feature of FQHE states. We
have performed a systematic search for incompressible states in
the $(N,2S)$ plane for spinless bosons with the pure delta
interaction. We have used the L\'anczos algorithm to obtain some
of the low-lying eigenvalues. Since incompressible states do not
break any symmetry, we expect that the corresponding spectrum has
a singlet $L=0$ ground state clearly separated by a gap from higher lying states.
Of course this gap should remain nonzero at the thermodynamic limit,
a fact that can be established only by careful finite-size scaling.
In Fig.(2) we display all states with a singlet ground state that
we have been able to study numerically. Some of these states are candidates
for an incompressible state. However this is not the case for all
the points in Fig.(2). When the number of bosons per state becomes quite large,
we find spectra that look like uncorrelated rotors with no sign of
gapped modes. This is the case for 2S=2,3 at all fillings, i.e.
all numbers of bosons.
Of course, the criteria mentioned above do not allow for an
unambiguous selection. One needs to find series of states
converging towards a thermodynamic limit. They should appear along
lines of fixed slope, i.e. filling factor, in the $(N,2S)$ plane.
In Fig.(2) we have plotted the lines corresponding to the FQHE states
$\nu =1/2, 2/3$ as well as the Moore-Read state for $\nu =1$ (to be discussed
below) and also the fraction $\nu=2$ from the unpaired hierarchy (this last one
does not converge to the thermodynamic limit, see sect. V).
\begin{figure}[!htbp]
\begin{center}\includegraphics[  width=3.25in,
  keepaspectratio]{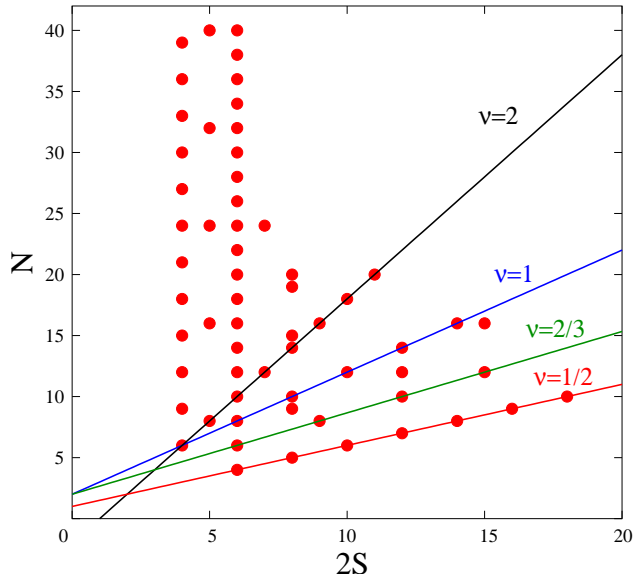}\end{center}
\caption{Candidates for incompressible states in the plane
(number of particles, flux). Dots indicate a $L=0$ ground state.
The lines show the flux-number of particles relationships of some FQHE states.}
\end{figure}

%%%%%%%%%%%%%%%%%%%%%%%%%%%%%%%%%%%%%%%%%%%%%%%%%%%%%%%%%%%%%%%%%%%%%%%%%%%%%%%%
\section{The principal sequence}
\subsection{The $\nu =1/2$ Laughlin state}
\begin{figure}[!htbp]
\begin{center}\includegraphics[  width=3.25in,
  keepaspectratio]{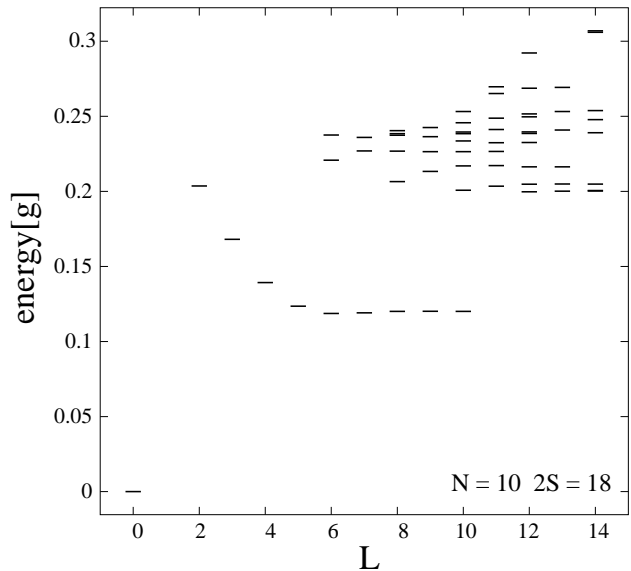}\end{center}
\caption{Energy spectrum for N=10 bosons at 2S=18 corresponding to
the filling $\nu =1/2$. The Laughlin state is the $L=0$ ground
state at zero energy. There is a clear collective mode branch that
extends up to $L=N=10$. The upper part of the spectrum has been
truncated for clarity but extends continuously.
 Energies are in units of $g$ and the horizontal axis
is total angular momentum.}
\label{LJFig}
\end{figure}

For $2S=2(N-1)$ we obtain the typical spectra from the
prototypical incompressible Laughlin state~: see Fig.(\ref{LJFig}). The
Laughlin state itself is the $L=0$ ground state. It is separated
by a clear gap from a set of states that are themselves separated
from the quasi-continuum of excited states. The well-defined
branch extend from $L=2$ up to $L=N$ and is a collective excitation
branch~\cite{Girvin85}. This is close to the case of fermions with Coulomb
interactions but there is one important difference~: here this
 branch has no clear minimum before it ends. In the
case of fermions, there is a minimum, the magnetoroton,
 that is attributed to the
incipient instability towards formation of a Wigner-crystal with a
well-defined lattice spacing, hence defining a typical wavevector.
In the case of a short-range interaction, we do not expect any
instability of this kind and hence no specific wavevector should
be singled out in the dispersion relation (on the sphere
wavevector correspond to angular momentum by $kR=L$). We have
evaluated~\cite{Regnault03} the gap between the ground state and
the collective mode by performing finite-size scaling up to N=10
bosons~: the result is $0.95(5)$g. This estimate does not vary
much when we use the extremal point $L=N$ of the collective mode
or any point in the flat part.

If we add one quantum of flux, then we create a single quasihole
with $L=N/2$ which is gapless (see section II). If we remove one
flux quantum, the quasielectron with also $L=N/2$ has a gap and no
exact result is known for its wavefunction. Similarly we have
computed the gap for quasielectrons for up to N=12 bosons and
extrapolated to the thermodynamic limit. The result is the same
gap as above for the collective mode. This is what we expect for a
quasihole-quasielectron bound state.

\subsection{The principal sequence}
\begin{figure}[!htbp]
\begin{center}\includegraphics[  width=3.25in,
  keepaspectratio]{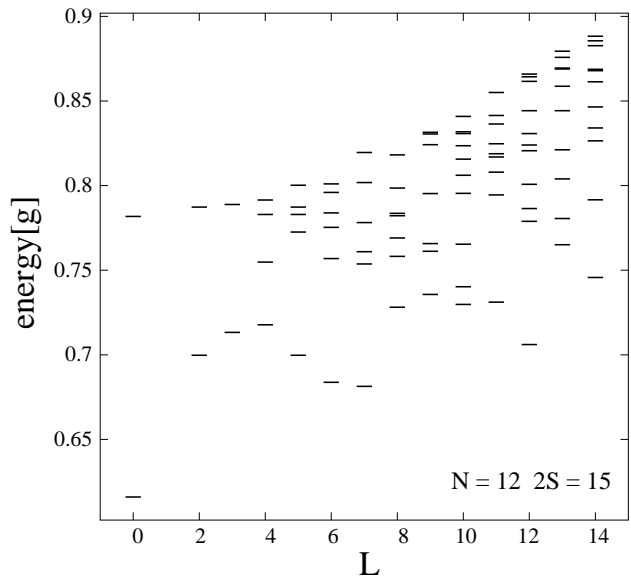}\end{center}
\caption{Energy spectrum for N=12 bosons at 2S=15 corresponding to the
filling $\nu =2/3$. Above the $L=0$ ground state there is
 a collective mode branch that extends up to $L=7$.}
\label{TwoThird}
\end{figure}

Starting from the state $\nu =1/2$ quasiparticle condensation into
successive Laughlin states generate a whole hierarchy of
incompressible states. In the case of fermions the most prominent
series are given by $\nu = n/(2n\pm 1)$. This so-called principal
sequence~\cite{Jain89} can be derived from the composite fermion
scheme~\cite{HeinonenBook} which is one of the variants of the
hierarchy theory. This can be easily translated for bosons.
Attaching a flux tube with one quantum of statistical flux
transmute the boson into a (composite) fermion. In a mean-field
picture the fictitious magnetic field compensates partially the
external applied field and the composite fermions then populate
Landau levels leading to incompressibility when an integer $p$ of
these levels are filled. With one flux tube per fermion the
filling of the original bosons is $\nu =p/(p\pm 1)$. On the sphere
the same reasoning leads to the relations~:
\begin{equation}\label{Hshift}
    2S=\frac{p\pm 1}{p}N \mp p -1 \quad\mathrm{for}\quad \nu=\frac{p}{p\pm 1}.
\end{equation}
From the accessible sizes, we have evidence for $\nu =2/3$
realized for N=4,6,8,10,12. The lower part of the spectrum for
N=12, 2S=15 is displayed in Fig.(\ref{TwoThird}). Since the filling
is now higher than 1/2 it is no longer possible that all bosons
avoid interactions, hence the ground state no longer has zero
energy. Above the isolated singlet ground state there is evidence
for a gapped collective mode which is not as clearly resolved as
in the case of the strongest fraction $\nu =1/2$. If we define the
gap as the energy difference between the ground state and the
first excited state irrespective of its angular momentum then an
extrapolation to the thermodynamic limit in 1/N leads to a nonzero
value of 0.05g. In the composite fermion scheme it is simple to
compute the  extend of the collective mode. The lowest energy
neutral excitation can be obtained by promoting a composite
fermion from the highest occupied $p^{th}$ Landau level to the
first empty $p+1^{th}$ level . Since the angular momenta are then
$L_{CF}=(N/2p)+(p/2)$ and $L_{h}=(N/2p)+(p/2)-1$
respectively for the fermion and the hole,
it follows that
the particle-hole state has maximal angular momentum
$L_{max}=(N/p)+p-1$. This is consistent with what we observe in
Fig.(4). For the Laughlin state with p=1, one has $L=N$ as the
maximum value which is the same result as considering the
collective mode as a quasihole-quasielectron pair. For all the
hierarchical descendants, we find the general trend that the
collective mode has apparently a downward curvature, as is also
seen in the case of fermions with Coulomb interactions.

For the fraction $\nu=2/3$ we have also studied the charged
excitations. From the composite fermion scheme we obtain that a
single quasihole (resp. quasielectron) is created by removal
(resp. addition) of a composite fermion, leading to the following
flux matching condition~:
\begin{equation}\label{S+}
\mathrm{quasiholes~:} \quad 2S=\frac{p\pm 1}{p}N \mp p -1 +\frac{1}{p}
\quad\mathrm{for}\quad \nu=\frac{p}{p\pm 1},
\end{equation}
\begin{equation}\label{S-}
 \mathrm{quasielectrons~:}\quad 2S=\frac{p\pm 1}{p}N \mp p -1 -\frac{1}{p}
 \quad\mathrm{for}\quad \nu=\frac{p}{p\pm 1}.
\end{equation}
Due to the factor $1/p$ it is not possible to create a
quasiparticle  without changing the number of bosons (except for
the primary Laughlin state $\nu =1/2$). As a consequence when
computing the gap to charged excitations one has to extrapolate
the ground state energy of the parent fluid to a number of bosons
at which it is not realized. For $\nu=2/3$ the gap estimate from
charged excitations converges to a value close to the gap of the
neutral excitations 0.05(1)g.

For the incompressible state at $\nu =3/4$ we have three
candidates at N=4,8,12. They all consistently display a gap and
the first branch of excited states also obeys the rule
$L_{max}=(N/p)+p-1$. There are too few points then to estimate the
gap. Similarly we find two consistent candidates for $\nu=4/5$ at
N=8 and 12. The same data set also points to the existence of at
least the fractions 4/3 and 5/4. Indeed, there is
aliasing~\cite{Ambrumenil89} of the the (N,2S) points~: a point
with a given value of N and 2S at fraction $\nu =p/(p+1)$ also
satisfies condition (\ref{Hshift}) for $\nu =
p^{\prime}/(p^{\prime}-1)$ if $pp^{\prime}=N$. We thus find
evidence for $\nu =4/3$ from N=8,12,16 and $\nu =5/4$ for N=5,10.

When the filling is less than 1/2,zero-energy states proliferate
and all other fractions are embedded in the angular momentum
descendants of the Laughlin state. This is a peculiarity of the
delta-function interaction and disappears if we change the
interaction pseudopotential. If we add a V$_2$ term then the most
prominent new fraction is $\nu =2/5$. This fraction does not belong to
the Jain principal sequence of states but appears if we consider
the Haldane bosonic construction of the hierarchy~: quasiholes on
top of the $\nu=1/2$ state can themselves condense into a
$\nu^\prime =1/2$ fluid leading to a 2/5 state (while it is
quasielectron condensation at $\nu^\prime =1/2$ that leads to the
2/3 state).

%%%%%%%%%%%%%%%%%%%%%%%%%%%%%%%%%%%%%%%%%%%%%%%%%%%%%%%%%%%%%%%%%%%%%%%%%%%%%%%%
\section{The $\nu =1$ state}
If we consider the composite fermion picture, a striking
prediction is that the effective magnetic field may be zero for
certain filling factor. For fermions this happens when $\nu =1/2$
which is the limiting point of the Jain sequence $n/(2n\pm 1)$ for
$n\rightarrow\infty$. In the case of the Coulomb interaction there
is evidence that some kind of Fermi sea of composite fermions is
present. This comes from studies of Chern-Simons
theory~\cite{Halperin93} and also from numerical
studies~\cite{Rezayi94} of spherical systems. Similarly for
bosons, the effective magnetic flux vanishes for $\nu=1$ the
limiting point of the sequence $n/(n\pm 1)$ for
$n\rightarrow\infty$. So some kind of renormalized Fermi liquid is
a likely candidate for the ground state at this special filling.
It has been also pointed out first for the Fermi case that the
effective Fermi sea may be unstable with respect to pairing of
composite fermions. An appealing wavefunction describing in a BCS
manner the pairing of the composite fermions is the so-called
Pfaffian wavefunction~\cite{Moore91,Greiter92,Milovanovic96}. On
the sphere it can be written~:
\begin{equation}\label{PfaffianEq}
    \Psi =\mathrm{Pf}\{\frac{1}{u_iv_j-u_jv_i}\}\prod_{i<j}(u_iv_j-u_jv_i)^q .
\end{equation}
The symbol Pf stands for the Pfaffian and $N$ should be even.
Given an antisymmetric $N\times N$
matrix $A_{ij}$ it is defined by~\cite{deGennesBook}~:
\begin{equation}\label{PfDef}
\mathrm{Pf}\{A_{ij}\}=\sum_\sigma \epsilon_\sigma
A_{\sigma(1)\sigma(2)}\ldots A_{\sigma(N-1)\sigma(N)},
\end{equation}
where $\sigma$ are permutations of the index with $N$ values. For
$q=1$ the wavefunction Eq.(\ref{PfaffianEq}) describes bosons at
filling factor $\nu =1$ while for $q=2$ it describes fermions at
$\nu =1/2$. The filling factor is $\nu=1/q$ and this wavefunction will fit
on the sphere if we have~:
\begin{equation}\label{fluxPf}
    2S=q(N-1)-1.
\end{equation}

The wavefunction in Eq.(\ref{PfaffianEq}) no longer vanishes when
two particles are at the same point because of the denominator. It
has thus no zero energy. However it still vanishes when three
particles coincides. Using this property one can construct a
special Hamiltonian for which it is the exact ground
state~\cite{Moore91}. Along a similar line, It has been also
noted~\cite{Cooper03} that a gas of atoms and molecules in the
condition of a Feshbach resonance may have also the Pfaffian as an
exact ground state.

\begin{figure}[!htbp]
\includegraphics[width=0.25\columnwidth, keepaspectratio]{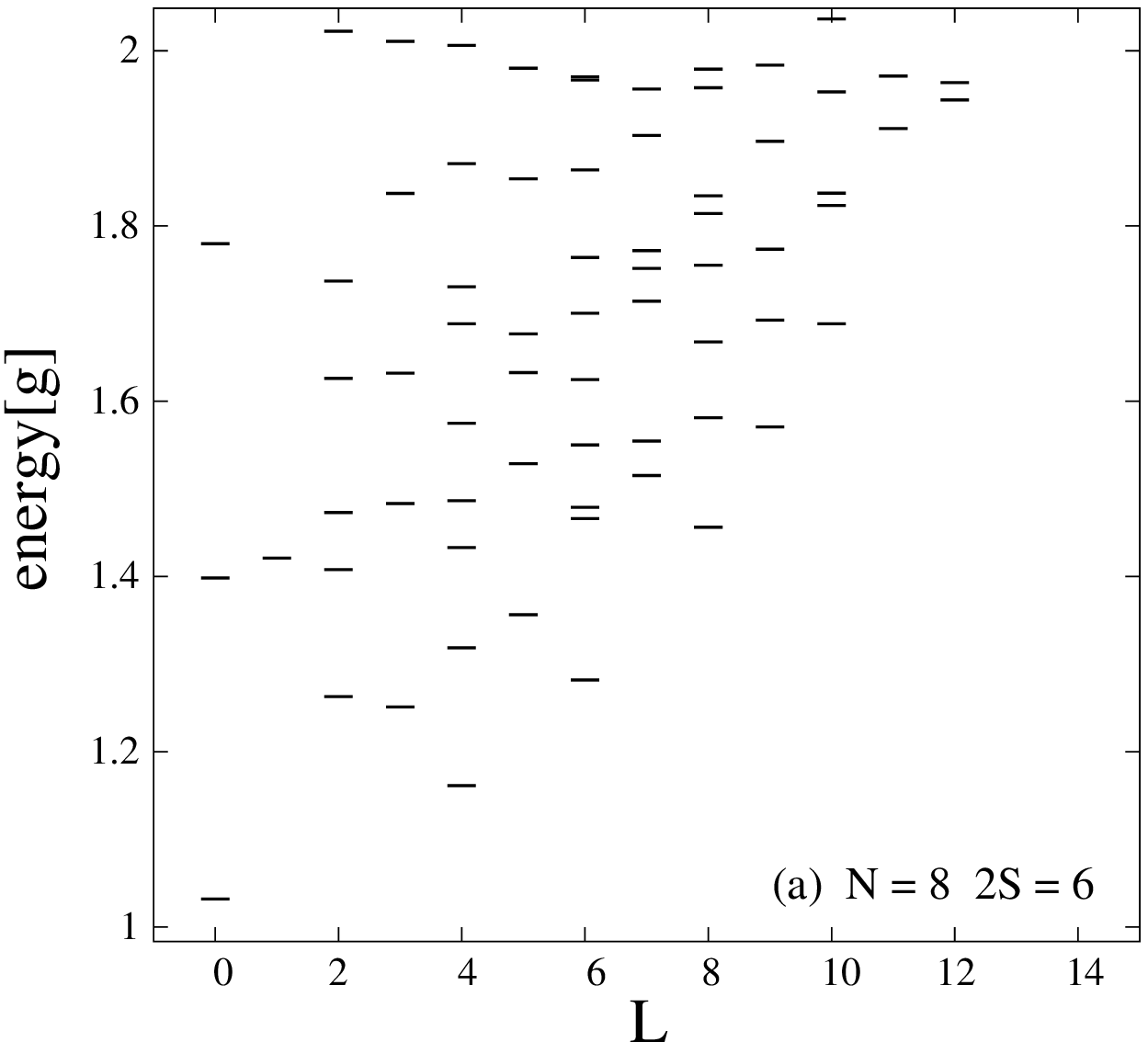}
\vspace{0pt}
\hspace{0.04\columnwidth}
\includegraphics[width=0.25\columnwidth, keepaspectratio]{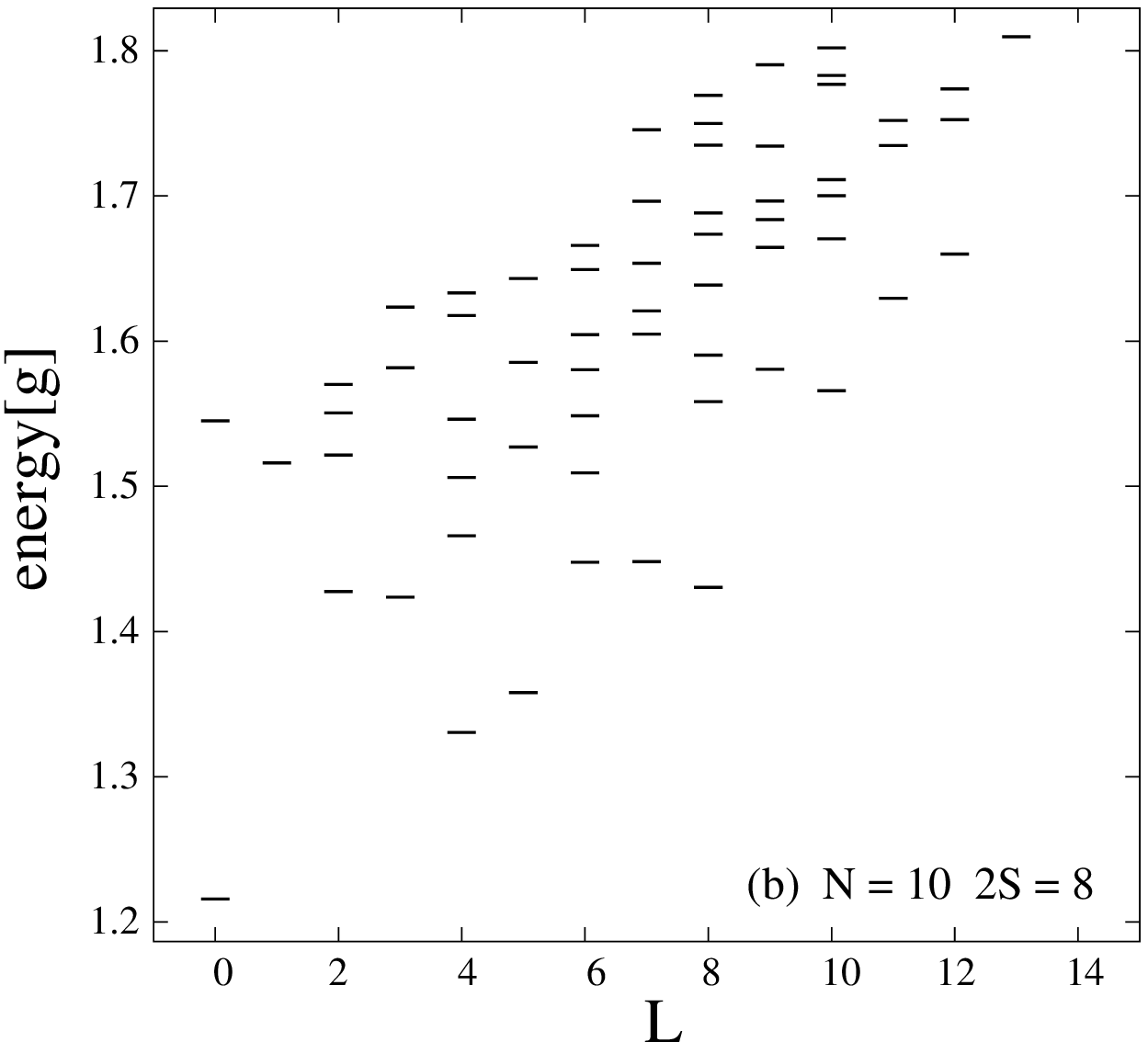}
\vspace{0pt}
\hspace{0.04\columnwidth}
\includegraphics[width=0.25\columnwidth, keepaspectratio]{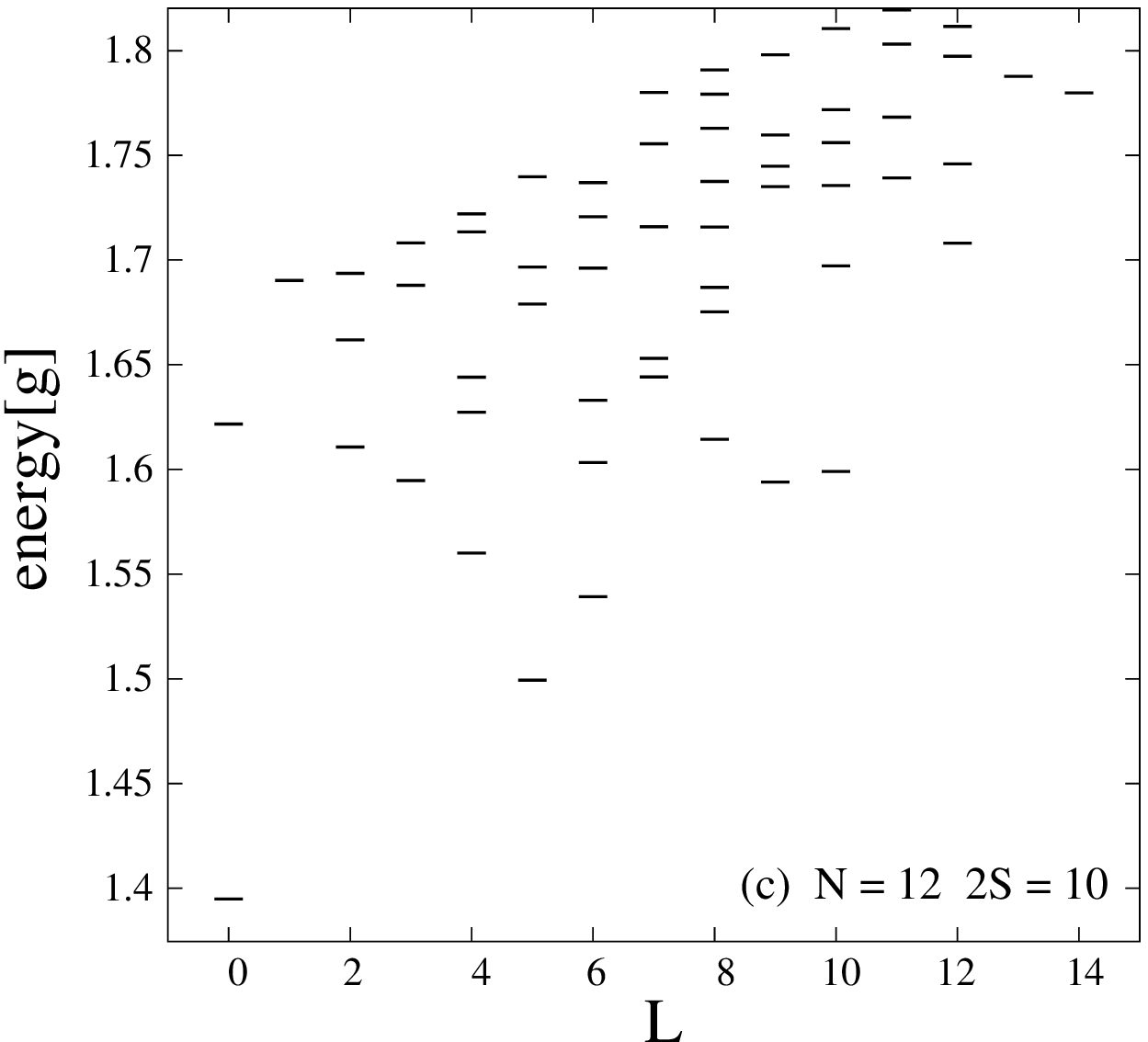}
\caption{Incompressible states related to the Pfaffian when the number
of particles is even~: (a) N=8, (b) N=10, (c) N=12.}
\label{EvenPf}
\end{figure}

\begin{figure}[!htbp]
\includegraphics[width=0.25\columnwidth, keepaspectratio]{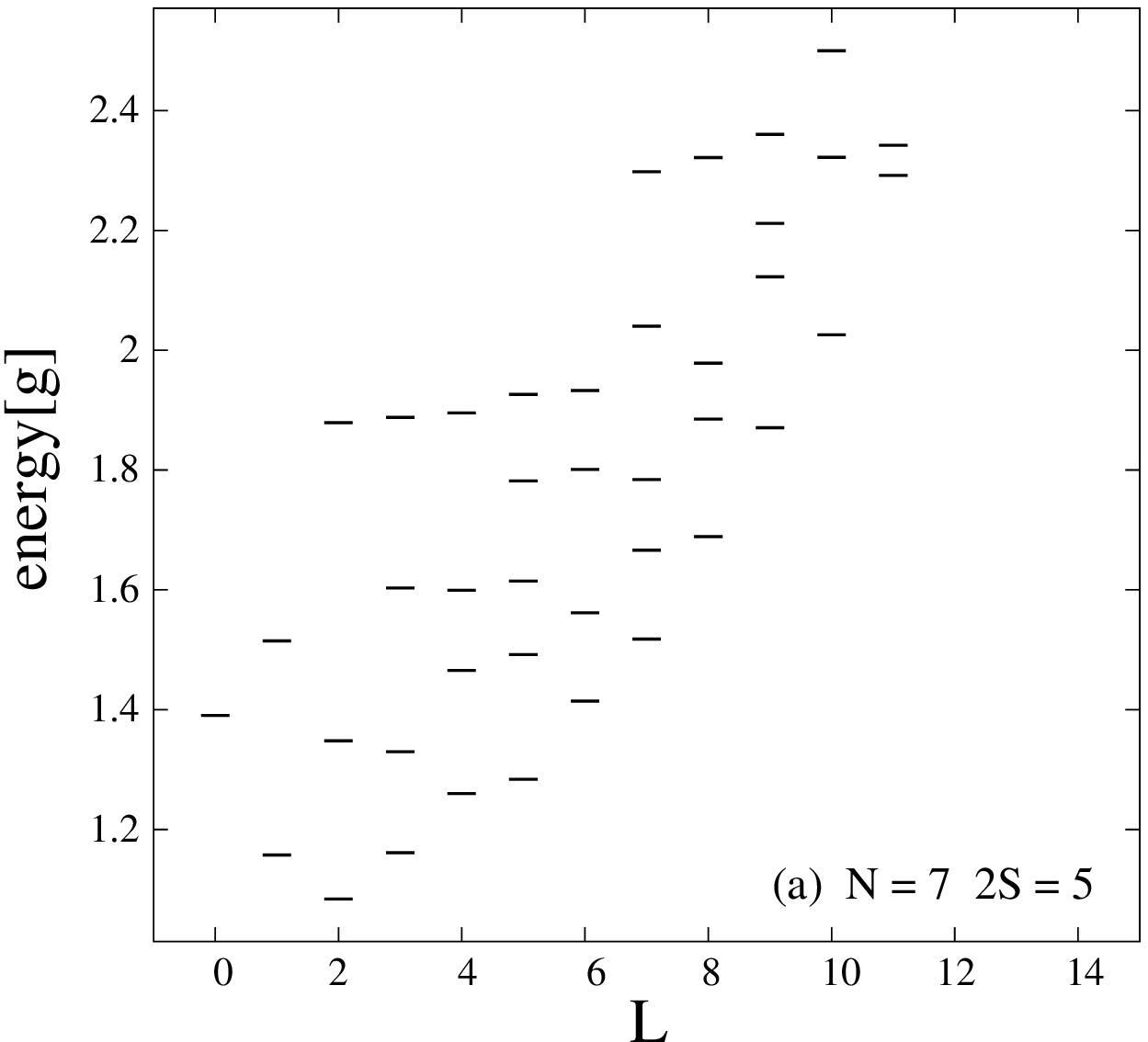}
\vspace{0pt}
\hspace{0.04\columnwidth}
\includegraphics[width=0.25\columnwidth, keepaspectratio]{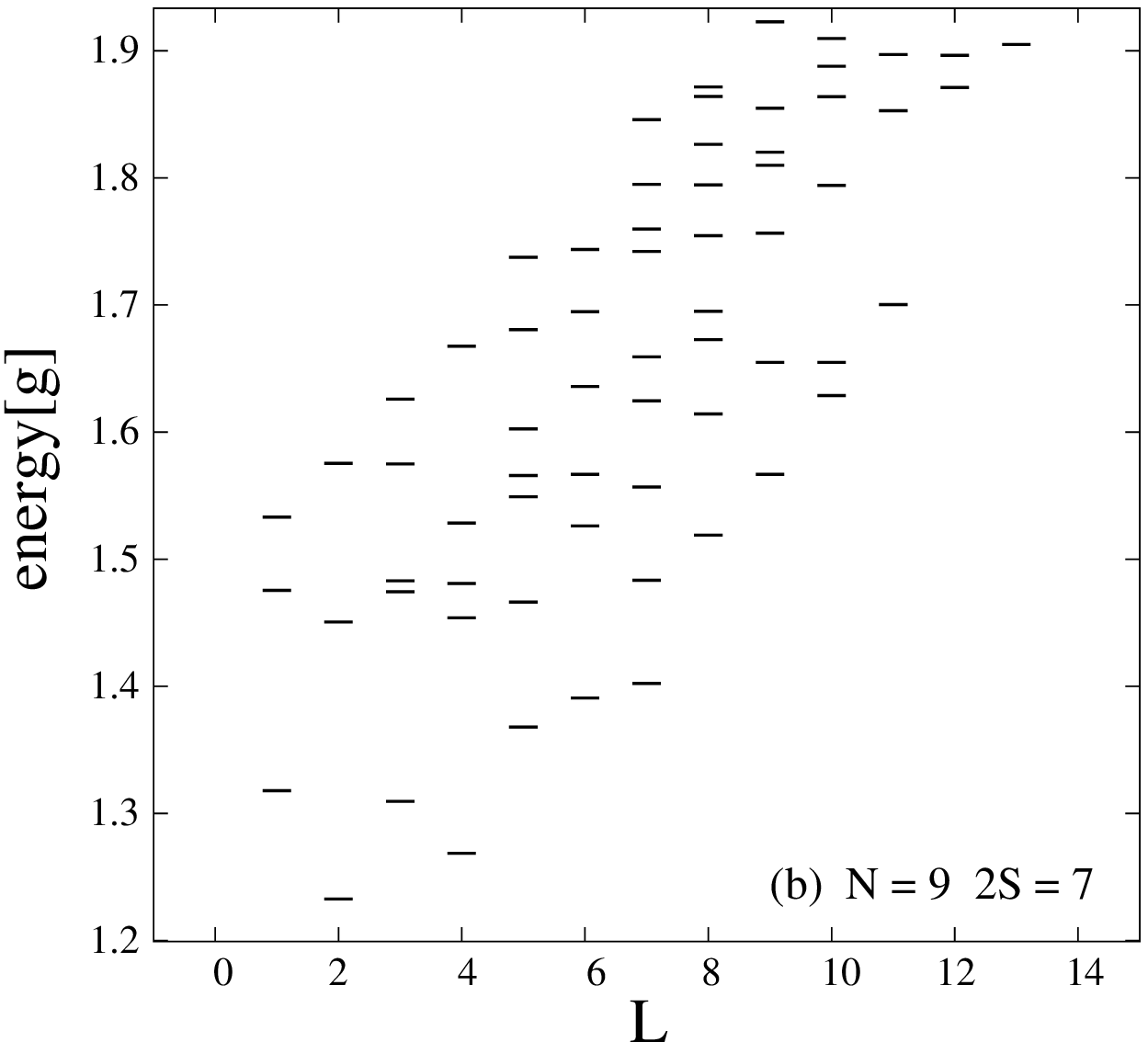}
\vspace{0pt}
\hspace{0.04\columnwidth}
\includegraphics[width=0.25\columnwidth, keepaspectratio]{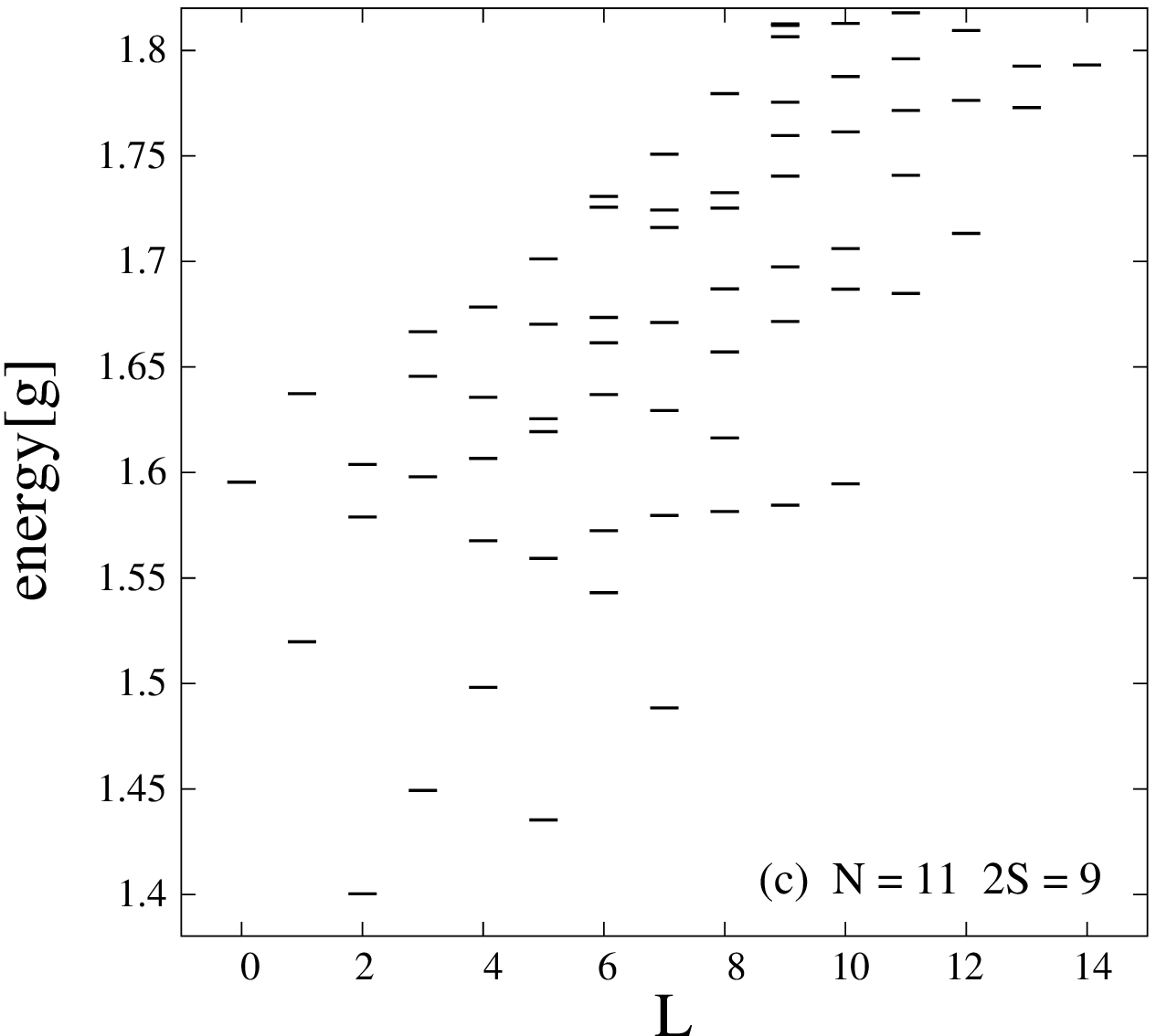}
\caption{States with one pair-breaking excitation when the number
of particles is odd~: (a) N=7, (b) N=9, (c) N=11.}
\label{OddPf}
\end{figure}

It has been proposed that the Pfaffian state describes the elusive
$\nu =5/2$ incompressible state observed in two-dimensional
electron gases~\cite{Morf98,Rezayi00}. One has to treat the LLL as
inert and then the lower spin branch of the first excited Landau
level which is at effective filling 1/2 may prefer the Pfaffian
paired state. There is some evidence that this happens from
numerical results on the sphere but this requires to tweak
somewhat the pseudopotential between electrons. In the Bose $\nu
=1$ case, studies in the torus geometry are consistent with the
Pfaffian state being the ground state~\cite{Cooper01}. There were
also early hints of this state~\cite{Canright89}. We observe in
the spherical geometry a strong series of incompressible
candidates satisfying $2S=N-2$ for N=4,6,8,10,12,14,16. Some of
them are displayed in Fig.(\ref{EvenPf}a,b,c). They have an
isolated singlet ground state and a well-defined branch of neutral
excitations  made of states separated by $\Delta L=1$. The gap
extrapolates to a nonzero value\cite{Regnault03} $\approx 0.1$g.
Even we use the less favorable fit, a quadratic fit starting from
the smallest size so including the maximum amount of downward
curvature in a 1/N plot, then the extrapolated value is still
nonzero. There are also neutral fermionic excitations above the
Pfaffian ground state. In the case of the bosonic problem, this
fermion is a composite fermion being one bare boson with one
additional flux tube~\cite{Read00}. From a paired state we obtain
such a state by adding one boson and adding also one extra flux
quantum. Since the paired state lies on the trajectory $2S=N-2$ we
are still on this line in the (N, 2S) plane but now with N
\textit{odd}. Consistently, we observe a strong even-odd effect on
the Pfaffian line. Typical spectra for odd number of particles are
given in Figs.(\ref{OddPf}a,b,c). They are gapless with a "hanging
chain" shape of states separated by $\Delta L=1$ as observed in
corresponding fermion systems~\cite{Greiter92}.

An intriguing feature of the Pfaffian physics is the occurrence of
unconventional excitations endowed with non-Abelian
statistics~\cite{Moore91}. This is especially interesting in the
case of the Bose systems in magneto-optical traps since  one may
eventually perform laser manipulations of quasiparticles as
suggested by Paredes et al.~\cite{Paredes01} to observe the
non-Abelian mixing of quasiparticles induced by a real-space
permutation of the positions. These excitations are created in
pairs by addition or removal of a single quantum of flux from the
magic condition Eq.(\ref{fluxPf}). The low-lying states should
form then a two-particle branch with a $\Delta L=2$ interval rule
as expected for indistinguishable particles.

We have also studied the stability of the Pfaffian state by
varying the interaction potential. This is done by adding more
pseudopotentials to the pure hard-core V$_0$ interaction. We use a
Hamiltonian of the form $H_0 + \lambda H_1$ where $H_0$ represents
the pseudopotential V$_0$ only and $H_1$ stands for all the even
pseudopotentials V$_m$ of the Coulomb interaction for $m\geq 2$.
The full Coulomb interaction is restored for $\lambda =1$. This
special point was studied in ref.(\cite{Xie91}).  For a N=6 boson
system on the sphere they found evidence for incompressibility
when $2S=4$, as expected for the Pfaffian state. We have computed
the gap above the Pfaffian in the range $0\leq\lambda \leq1$. We find
that the finite size law of the gap barely changes as we go to a
long-range potential. In all cases the extrapolation is consistent
with a nonzero gap. So it is likely that this paired
state is robust. When $\lambda\neq 0$, the states for $\nu\leq
1/2$ are no longer zero-energy states and we find that the most
prominent fraction that emerges is $\nu =2/5$.
%%%%%%%%%%%%%%%%%%%%%%%%%%%%%%%%%%%%%%%%%%%%%%%%%%%%%%%%%%%%%%%%%%%%%%%%%%%%%%%%
\section{Incompressibility at $\nu > 1$}
\begin{figure}[!htbp]
\begin{center}\includegraphics[  width=2.25in,
  keepaspectratio]{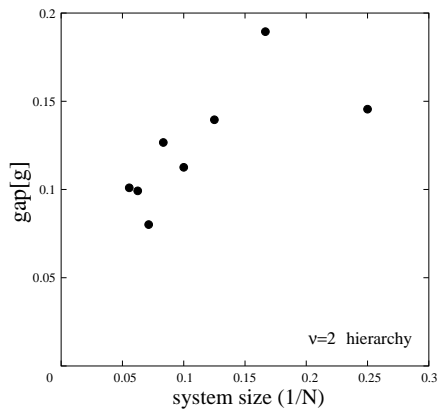}\end{center}
\caption{The gaps of the candidate states for a hierarchical $\nu =2$ fraction.}
\label{H2}
\end{figure}

When the filling factor is large then it is likely that there will
be a quantum phase transition towards the Abrikosov lattice of
vortices already observed in trapped gases. Initial
estimates~\cite{Cooper01} from exact diagonalizations in the torus
geometry point to a critical filling $\nu_c \approx 6$ for the
transition. A study using a Lindemann melting criterion for the
vortex lattice also gives a similar result~\cite{Sinova02}: $\nu_c
\approx 8$. It is thus of interest to understand the special
region for fillings larger than one. There are hierarchy states
belonging to the principal sequence~: this is the case for $\nu
=3/2$ and $\nu =2$ the termination point of the $p/(p-1)$ series
of states. For $\nu =3/2$ we have only three candidates available
N=6,9,12 and N=12 is aliased with the Pfaffian. While the
corresponding spectra look incompressible, it is not possible to
draw any firm conclusion concerning their survival at the
thermodynamic limit. For $\nu =2$ on the contrary, the values
N=6,8,10,12,14,16,18 can be studied. In this series the
neutral gaps as a
function of 1/N are displayed in Fig.(\ref{H2}). There is no
tendency to convergence. If we try to investigate charged
excitations on top
of the $\nu =2$ state, then we find only compressible
states but no clear quasiparticles.

\begin{figure}[!htbp]
\includegraphics[width=0.25\columnwidth, keepaspectratio]{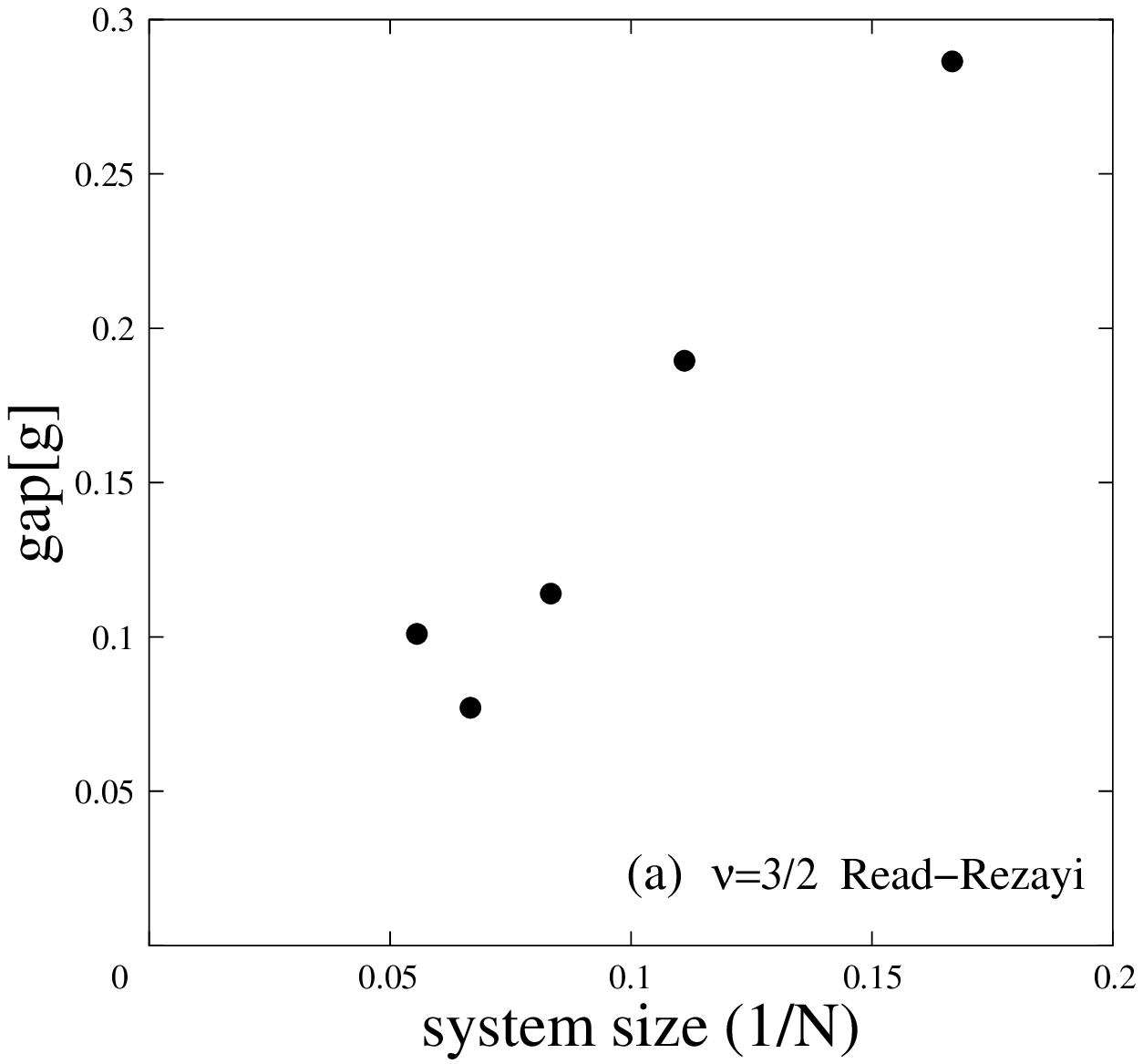}
\vspace{0pt}
\hspace{0.04\columnwidth}
\includegraphics[width=0.25\columnwidth, keepaspectratio]{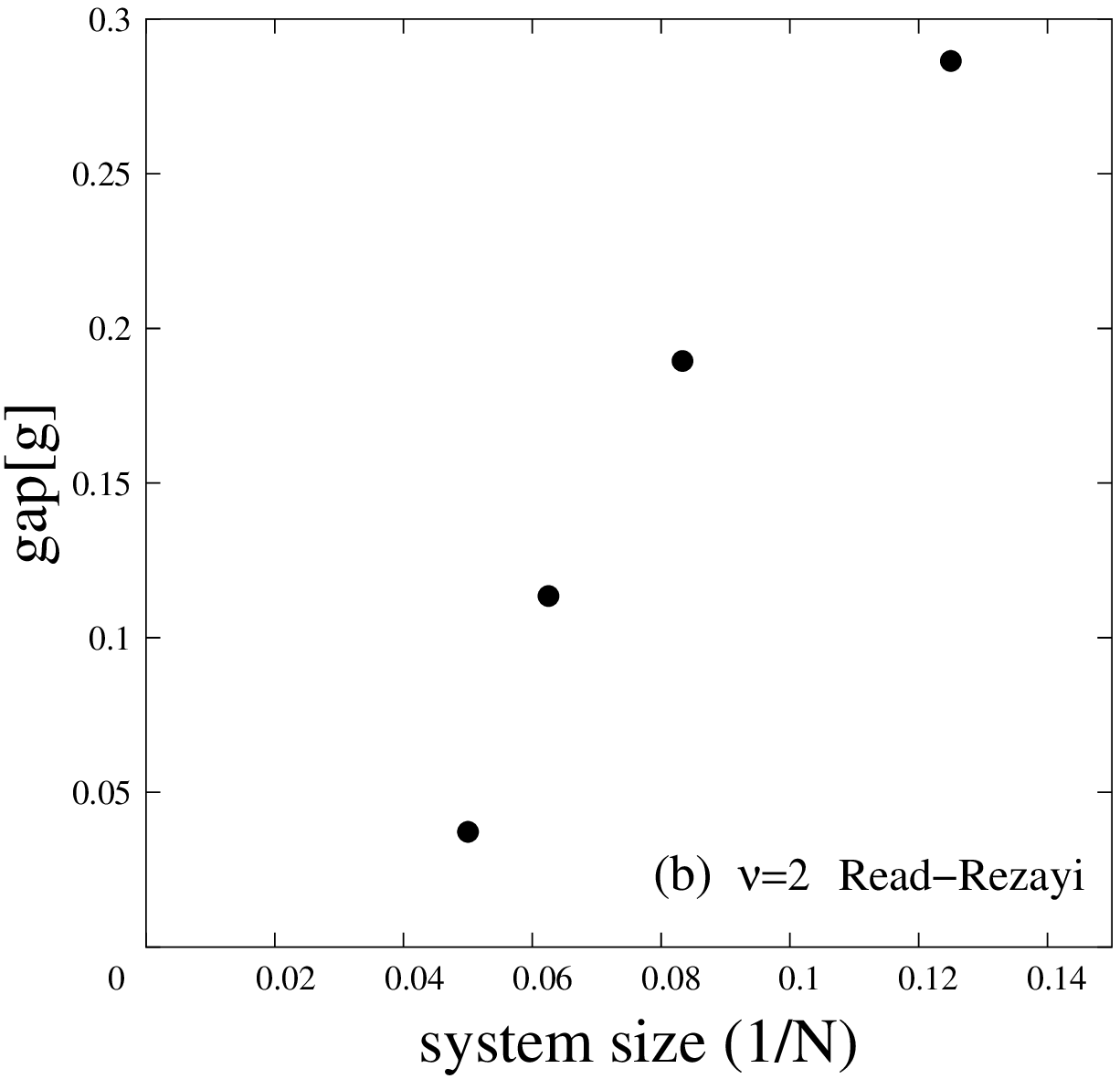}
\caption{Gaps for the possible Read-Rezayi states on the sphere~: in (a) the
fraction $\nu =3/2$ is displayed (b) the fraction $\nu =2$.}
\label{RRGaps}
\end{figure}

This is consistent  with the proposal that in this regime of filling factors
the system
has ground states belonging to a family of states containing clusters of $k$
particles, the Read-Rezayi states~\cite{ReadRezayi,Read00}.
The corresponding wavefunctions
are constructed by symmetrizing products of $k$ Laughlin states. For $k=1$ this
gives simply the Laughlin state Eq.(\ref{LJ}) and for $k=2$ the Pfaffian state
Eq.(\ref{PfaffianEq}). The filling factor is given by $\nu =k/2$ and on the sphere
they require the special flux $2S=(2/k)N-2$. For $\nu =3/2$ we are
able to study N=6,9,12,15 and for $\nu =2$ N=8,12,16. Again the spectra shows
the features associated with FQHE states. We have obtained the neutral gap
of these states. However it is difficult to draw any firm
conclusion concerning the limit of large system size because the gaps are
badly behaved~: see Figs.(\ref{RRGaps}a,b).

Finally we note that the Read-Rezayi states have several quasihole states~:
adding one flux quantum to the fiducial state leads to the production
of $k$ quasiholes. They should appear as a set of low-lying levels with an
easily computable degeneracy. This degeneracy would be exact for the model
Hamiltonian with ($k+1$)-body forces of Read and Rezayi. For the system sizes of
this work, this does not appear clearly but is not surprising. It is likely
that these fractions have large correlation lengths and the finite size
effects are still too large.

%%%%%%%%%%%%%%%%%%%%%%%%%%%%%%%%%%%%%%%%%%%%%%%%%%%%%%%%%%%%%%%%%%%%%%%%%%%%%%%%
\section{Conclusions}
In this paper we have studied the occurrence of quantum Hall
fractions of spinless bosons in a rotating trap in the lowest
Landau level. We have used exact diagonalization techniques in the
spherical geometry. We gave evidence for the appearance of the
bosonic analog of Jain principal sequence of FQHE fractions $\nu
=p/(p\pm 1)$ instead of $\nu =p/(2p\pm 1)$ for fermions. This is
exactly what is expected from standard arguments based on
Fermi-Bose correspondence and Chern-Simons theories, albeit not
rigorously established. In the spectra of these fractions, we have
shown that there exists states corresponding to the expected
collective density mode of the FQHE states. The angular momentum
extent of the collective mode is in agreement with the standard
physics of FQHE fractions. There is so far no evidence for
fractions outside the principal sequence. This is not so if we
modify the potential between bosons~: then notably the fraction
$\nu =2/5$ appears clearly. This principal sequence cannot
describe the physics of the bosonic fluid for fillings larger than
one. In fact its two members $\nu =2$ as well as $\nu =3/2$ do not
show any indication of convergence towards the thermodynamic
limit.

For complete filling of the lowest Landau level, we have shown
that the potential Fermi sea of composite fermions is destroyed and that
a paired state, the Pfaffian state is likely to be the ground state.
There is a series of incompressible states that has exactly the shift
predicted by the Pfaffian wavefunction. The gap extrapolates smoothly to
a nonzero value and in addition there is good evidence that half-flux
excitations typical of this paired state are present.
At the points in the (N,2S) plane predicted by Read and Rezayi we
find also FQHE states for fillings larger than one. In this case
their ultimate fate in the thermodynamic limit is not clearly established
from the system sizes we have been able to treat.
This remains a subject for future studies.

%%%%%%%%%%%%%%%%%%%%%%%%%%%%%%%%%%%%%%%%%%%%%%%%%%%%%%%%%%%%%%%%%%%%%%%%%%%%%%%%

\begin{acknowledgments}
We thank Jean Dalibard for fruitful discussions.
The numerical calculations have been performed thanks to a
computer time allocation of IDRIS-CNRS.
\end{acknowledgments}

%%%%%%%%%%%%%%%%%%%%%%%%%%%%%%%%%%%%%%%%%%%%%%%%%%%%%%%%%%%%%%%%%%%%%%%%%%%%%%%%

\end{document}